\pdfoutput=1

\documentclass{sig-alternate}
\usepackage{amsmath}
\usepackage{graphicx}
\usepackage{url}
\usepackage{booktabs}
\usepackage{amssymb}
\usepackage{latexsym}	
\usepackage{array}		
\usepackage{multirow}
\usepackage{subfigure}
\usepackage{algorithm}
\usepackage{algpseudocode}
\usepackage{threeparttable}
\usepackage{paralist}   
\usepackage{xspace}     
\usepackage{color}

\newcommand{\figurename}{Figure\xspace} 

\newcommand{\toolname}{ECVDetector\xspace}
\newcommand{\toolnameno}{ECVDetector\xspace}

\newcommand{\DIRECT}{\texttt{VS\_Direct}\xspace}
\newcommand{\DIRECTno}{\texttt{VS\_Direct}\xspace}
\newcommand{\DIRECTBy}{\texttt{VS\_DirectByParam}\xspace}
\newcommand{\DIRECTByno}{\texttt{VS\_DirectByParam}\xspace}
\newcommand{\SEND}{\texttt{VS\_Input}\xspace}
\newcommand{\SENDno}{\texttt{VS\_Input}\xspace}
\newcommand{\PUBLIC}{\texttt{VS\_Public}\xspace}
\newcommand{\PUBLICno}{\texttt{VS\_Public}\xspace}

\newcommand{\SOURCE}{\texttt{VSource}\xspace}
\newcommand{\SOURCEno}{\texttt{VSource}\xspace}

\begin{document}
%
\toappear{} 

\title{A Sink-driven Approach to Detecting\\ Exposed Component Vulnerabilities in Android Apps}

\author{
%
%
\alignauthor
Daoyuan Wu, Xiapu Luo, and Rocky K. C. Chang\\
       \affaddr{Department of Computing}\\
       \affaddr{The Hong Kong Polytechnic University}\\
       \email{\{csdwu, csxluo, csrchang\}@comp.polyu.edu.hk}
}

\maketitle

\begin{abstract}

Android apps could expose their components for cooperating with other apps.
This convenience, however, makes apps susceptible to the \textit{exposed component vulnerability} (ECV), in which a dangerous API (commonly known as \textit{sink}) inside its component can be triggered by other (malicious) apps.
In the prior works, detecting these ECVs use a set of sinks pertaining to the ECVs under detection.
In this paper, we argue that a more comprehensive and effective approach should start by a systematic selection and classification of \textit{vulnerability-specific sinks} (VSinks).
The set of VSinks is much larger than those used in the previous works.
Based on these VSinks, our sink-driven approach can detect different kinds of ECVs in an app in two steps.
First, VSinks and their categories are identified through a typical forward reachability analysis.
Second, based on each VSink's category, a corresponding detection method is used to identify the ECV via a customized backward dataflow analysis.
We also design a semi-auto guided analysis and validation capability for system-only broadcast checking to remove some false positives.
We implement our sink-driven approach in a tool called \toolname and evaluate it with the top 1K Android apps.
Using \toolname we successfully identify a total of 49 vulnerable apps across all four ECV categories we have defined.
To our knowledge, most of them are previously undisclosed, such as the very popular Go SMS Pro and Clean Master.
Moreover, the performance of \toolname is high, requiring only 9.257 seconds on average to process each component.

\end{abstract}

\section{Introduction}
\label{sec:intro}

To ease and accelerate the development of apps, Android takes a modular programming paradigm that empowers developers to focus on essential building blocks (i.e., components \cite{component} in Android terminology).
Moreover, Android apps could expose their components for cooperating with other apps.
For example, both Facebook and Twitter apps leverage an existing photo-capturing component exposed by a camera app by simply sending a request to it.
The photo-capturing component in this case is an \textit{exposed component} that serves external requests from other apps.

In return for this convenience, exposed components, if not well designed, might run into security risks.
In fact, vulnerabilities might exist if a dangerous API inside exposed components can be triggered by other (malicious) apps.
We refer to this class of vulnerabilities as \textit{exposed component vulnerability} (ECV), and the dangerous APIs in ECVs are the sinks of potential attack flows.
Usually, ECVs could be exploited by an attacker to perform dangerous operations by simply sending crafted inputs from a regular app to a victim app, both installed on the same phone.

Several methods for detecting specific ECVs \cite{Woodpecker12, DroidChecker12, CHEX12, ContentScope13, CustomRom13, IntentFuzzer14} have been proposed in the past.
In all of these works, detecting these ECVs use a set of sinks pertaining to the ECVs under detection.
Specifically, Woodpecker \cite{Woodpecker12}, DroidChecker \cite{DroidChecker12}, and the very recent IntentFuzzer \cite{IntentFuzzer14} are designed to detect \textit{permission leakage} in Android apps,
and they focus on a specific kind of sinks that would directly leak permissions once victim apps are exploited.
A recent work, SEFA \cite{CustomRom13}, further considers some database-related sinks that are aimed by ContentScope \cite{ContentScope13} for a special kind of components (i.e., Content Provider).
Finally, CHEX \cite{CHEX12} discovers potential vulnerabilities related to another kind of sinks, which are the data sinks that might cause unauthorized read or write operations on sensitive resources.

In this paper, we argue that a more comprehensive and effective approach should start by a systematic selection and classification of sinks.
Note that in the context of ECV detection, sinks should be vulnerability-specific (i.e., vulnerability-specific sinks, or VSinks) in contrast to the general data sinks for privacy leak detection \cite{SuSi13}.
This approach will help resolve two major issues in the previous detection methods.
First, the set of sinks obtained from our approach is much more comprehensive.
It will therefore help the previous methods to discover new ECVs.
Another and also more important issue is that the prior methods are tightly coupled with individual analysis requirements of their selected sinks.
They therefore cannot collaborate with one another to form a more general detection method.
Our approach, on the other hand, breaks this coupling by admitting different kinds of sinks and categorizing them for different analysis methods.


Using this sink-driven approach, we adopt a systematic strategy to select VSinks and classify them into multiple categories according to their different analysis requirements.
In this strategy, we combine multiple metrics (e.g., permission semantics and API names) to systematically define rules.
These rules are made according to a simple, but practical, rule syntax.
We further write a rule interpreter to automatically select and classify VSinks according to the defined rules.


Based on the categorized VSinks, our sink-driven approach can detect different kinds of ECVs in an app in two steps.
First, VSinks and their categories are identified through a typical forward reachability analysis.
We employ an iterative intra-procedural algorithm with flow sensitivity to perform this reachability analysis.
Second, based on each VSink's category, a corresponding detection method is used to identify the ECV via a customized backward dataflow analysis.
The backward, instead of prior forward, dataflow analysis is chosen to adapt more categories of sinks and organically cooperate with the forward analysis.
Furthermore, we design a semi-auto guided analysis and validation capability for system-only broadcast checking for removing some false positives.

In summary, this paper makes the following contributions:

\begin{itemize}
  \item \textbf{Methodology.} We propose a new sink-driven approach to systematically tackling the ECV detection problem (\S \ref{sec:method}).
  This approach includes a systematic strategy for VSink selection and classification, a general detection method to identify all categories of potential ECVs, and semi-auto guided analysis for excluding some sink-specific false positives.

\item \textbf{Tool and dataset.} We implement our sink-driven approach in a tool called \toolname (\S \ref{sec:implement}).
  We also design three analysis enhancements in \toolname, and the major one is that \toolname can validate broadcast checking, a capability that could significantly reduce false positives.
  Moreover, \toolname identifies a total of 372 VSinks across four categories, as well as 183 data source APIs.
  We are going to release this dataset to the Android research community.

\item \textbf{Evaluation and results.} We evaluate \toolname with the top 1K Android apps from Google Play (\S \ref{sec:evaluate}).
  In total, we identify 49 vulnerable apps across all the four ECV categories.
  To our knowledge, most of them are previously undisclosed, such as the very popular Go SMS Pro and Clean Master.
  Moreover, the performance of \toolname is high, requiring only 9.257 seconds on average to process each component.
\end{itemize}


\section{Background}
\label{sec:background}


\subsection{Exposed Component}
\label{sec:exposed}

Exposed components are a subset of Android-defined \textit{exported components} (i.e., components that other apps could access).
Some exported components have reliable permissions to protect them, but the exposed components in our threat model are fully exposed to other zero-permission apps or apps with only normal level permissions.
This threat model is also adopted in previous related work \cite{Woodpecker12, ContentScope13}.
We detect exposed components according to the following rule:

\textbf{RULE 1 (exposed component determination).}
A component is considered as an exposed component when it satisfies both two conditions:
\emph{\\
\indent C1: It must be enabled so that it could be successfully instantiated by the system; AND\\
\indent C2: It must be explicitly or implicitly exported so that other app could access it without permission or only with \texttt{normal} level permission.
}

Each Android app contains an \texttt{AndroidManifest.xml} file, which defines a set of component attributes.
Therefore, the rule for the exposed component determination is to inspect corresponding attributes.
Using \textit{C1}, we can exclude those useless components (i.e., those who set \texttt{enabled} attribute as false), such as the already deprecated components.
For \textit{C2}, the \textit{explicitly exported component} are those with their \texttt{exported} attribute set to true.
The \textit{implicitly exported components} are by default exported by Android convention.
For example, Intent-based components with \texttt{intent-filter} tag and Content Provider components without \texttt{exported} attribute are implicitly exported (Note that this default Provider convention is disabled since Android 4.2, but all other lower Android versions have to be compatible with this convention.).

\subsection{Problem Statement}
\label{sec:problem}

Exposed Component Vulnerability, ECV, is one kind of the classic confused deputy vulnerabilities \cite{Confused88}.
The exposed component mechanism in Android allows other apps to send a request or input to the victim component.
However, if the victim component cannot differentiate whether the request is from trusted parties or not and blindly execute its own code for finishing the request, then it becomes as a confused deputy.
Furthermore, if the triggered ``deputy'' would execute some security-related operations, then an ECV exists.
In this case, the victim component is a stepping stone for attackers and also the direct executor of attack behaviors.
A model of such attack scenario could be found in \cite{IPCInspection11}.
Essentially, the ECV is originated from insecure IPC communication in Android (mainly delivered by \textit{Intent} this IPC object), and a related problem is the attack on unauthorized Intent receipt \cite{ComDroid11}.
However, this issue is mainly due to the ambiguity during the resolution of Intent messages, thus not related to the ECV.

The impact of ECVs largely depends on the triggered sinks and the working mechanism of victim app itself.
Some ECVs are serious (such as forcing the victim app to send a SMS), while others could be even treated as bugs (e.g., starting a private Activity for attackers).
It is nearly impossible to give a clear boundary between such vulnerabilities and bugs.
In this paper, we generally consider these two kinds as vulnerabilities:
(1) those cause security or privacy risks to the user or phone,
and (2) those have security impact to internal status of victim apps.
Moreover, issues in exposed Activity components that require user interaction are not included in our threat model, since they are not stealthy and usually not recognized by vendors.


Our scope of ECV detection in this paper is the same as two previous related work \cite{Woodpecker12, CHEX12} that use the sink-based flow analysis.
However, only detecting attack surfaces and giving warnings (one focus of \cite{ComDroid11, ICCEpicc13}) is not enough for our goal.
Instead, in this work we try to discover the real exposed component vulnerabilities.
On the other hand, the complicated attacks on unauthorized origin crossing \cite{CrossOrigin13} is out of our scope, because it relies on too much browser-specific domain knowledge.
Moreover, two technical issues (Java reflection and native code), are out of our paper scope.
This also implies that we only need to select sinks from documented APIs (i.e., those could be found in Android SDK).

\subsection{VSink and its Taxonomy}
\label{sec:evolution}

VSink, short for Vulnerability-specific Sink, is a sink API where an attack flow could go to and related to ECVs.
In this paper, we consider VSinks mainly from these two channels:
\emph{\\
\indent C1: APIs that will cause security risks to permission-protected or privileged resources; OR\\
\indent C2: APIs that will make security impact on internal status of apps.\\
}
VSinks can be further classified into four categories according to their different analysis requirements.
We introduce these four categories as follows.



\textbf{\DIRECT.}
This category of VSinks is usually related to privileged resources, and they can be directly used to launch an attack.
For instance, the \texttt{removeAccount} is a \DIRECT sink.
Another example is \texttt{SmsManager.sendTextMessage()}, which can send a SMS message to outside without the attention of user.
Analyzing \DIRECT sinks is relatively straightforward, usually based on a path reachability analysis.
A major difference between VSinks and sinks in other Android works is that not all permission-required APIs will be considered as VSinks, only the vulnerability-related ones.
For example, \texttt{WAKE\_LOCK} and \texttt{VIBRATE} APIs are excluded.

\textbf{\DIRECTBy.}
This category of VSinks is similar to \DIRECTno, but they rely on the incoming parameters to exhibit different attack behaviors.
Different parameters could cause different attack consequences.
For instance, \texttt{ContentResolver.delete(Uri)} could exhibit different attack behaviors with different URI parameters, such as ``content://sms'' for deleting SMS.
Analyzing such ECVs is more complex than \DIRECTno, because a chain of variable dependences must be investigated, so that we can obtain the value of parameters.
Note that this category of VSinks has not \textit{generally} been considered in prior works (e.g., CHEX \cite{CHEX12} only consider some similar APIs when external Intent can control their parameters, while such relationship between Intent and parameters in our \DIRECTBy is not necessarily required).

\textbf{\SEND.}
This category of VSinks mainly fulfills the goal of misusing privileged resources, and this kind of misuse relies on attack inputs to flow into \SEND sinks.
Network-related sinks are the typical examples of \SENDno, such as \texttt{HttpClient.execute()}.
Once attack inputs flow into them, they could be exploited to misuse protected Internet resources.
Besides network-related \SENDno, APIs, such as \texttt{startService()} and \texttt{startActivity()}, also belong to this category.
Note that although \SEND sinks usually contain parameters, this is not necessarily required.
For instance, inputs flow into the caller object of \texttt{HttpURLConnection.connect()} will also cause \SEND ECVs.

\textbf{\PUBLIC.}
VSink APIs in this category are those which might not directly transmit privileged resources to attackers but will make them public to other apps.
An attacker could then leverage a local app to steal privileged resources outputted from these \PUBLIC APIs.
One typical kind of \PUBLIC sinks is the output methods defined in the \texttt{android.util.Log} class, such as debug-level method \texttt{d(String tag, String msg)}.
Vulnerable components might put privileged resources (e.g., GPS locations) to these output methods, and an attacker could collect these log outputs in runtime.

Besides VSinks, data source APIs are also related to specific ECV detection (e.g., \PUBLIC ECVs).
In this paper, we consider the data source APIs that can read permission-protected resources and call them \SOURCEno, such as the \texttt{BLUETOOTH} APIs.
\section{\toolname Design}
\label{sec:method}

\begin{figure}[ht!]
\begin{center}
\includegraphics[width=0.46\textwidth]{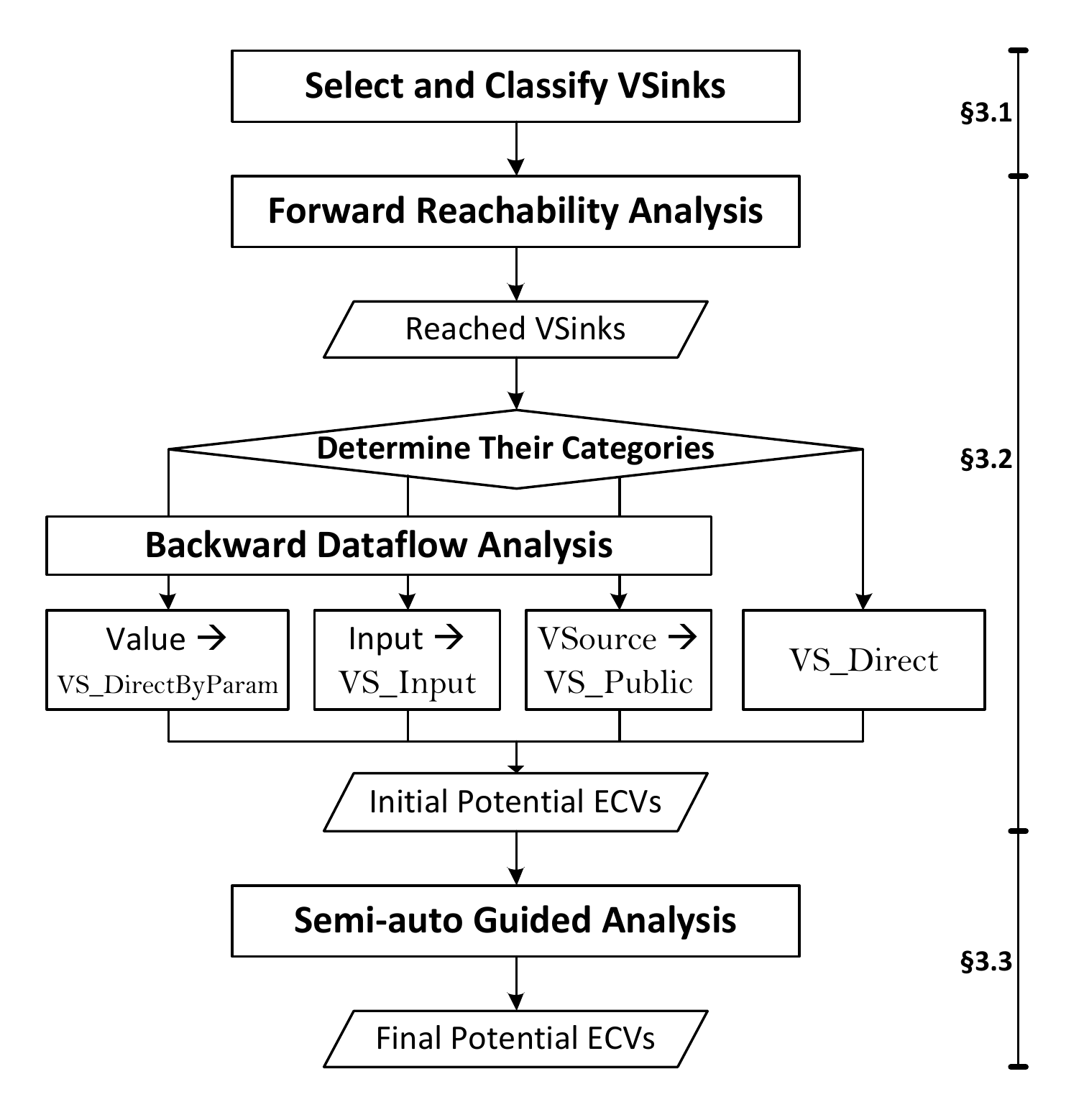}
  \vspace{-5ex}
\end{center}
\caption{The overall work flow of \toolnameno.}
\label{fig:workflow}
\end{figure}

In this section, we present our design of \toolnameno, which implements the sink-driven approach to systematically tackling the ECV detection problem.
As shown in \figurename \ref{fig:workflow}, \toolname first selects and classifies VSinks.
Then based on these VSinks, we propose a general detection method to identify all categories of potential ECVs.
This method organically combines forward reachability and backward dataflow analysis, and drives them by the characterized VSinks.
In particular, we take the backward, instead of previous forward, dataflow analysis for adapting more categories of sinks, such as the \DIRECTBy category.

More specifically, the first step produces a number of categorized VSinks, which are the dataset for the subsequent analysis.
Then \toolname employs a core module with a typical forward reachability analysis to tackle the common analysis task (i.e., identifying all reachable VSink calls from attack entry points), and further leverages a customized backward dataflow analysis to handle different categories of VSinks in several dedicated modules.
As the last step, a semi-auto guided analysis is conducted for excluding some sink-specific false positives.
In the following three sections, we will discuss each critical analysis phase in more detail.

\subsection{VSink Selection and Classification}
\label{sec:sink}

%
%

We propose a systematic strategy for VSink selection and classification.
The basic idea of this strategy is that we combine multiple metrics (e.g., permission semantics and API names) to systematically define rules.
These rules are made according to a simple, but practical, rule syntax.
We further write a rule interpreter to automatically select and classify VSinks according to the defined rules.


Most of VSinks are selected from permission-protected or privileged APIs, according to the \textit{C1} channel in \S \ref{sec:evolution}.
However, Android does not provide a complete and accurate mapping for permissions and their corresponding API calls.
In fact, permission information provided by Android developer document is limited and might contain errors \cite{Stowaway11}.
However, two great works \cite{Stowaway11, PScout12} have attempted to address this limitation.
Specifically, Stowaway \cite{Stowaway11} constructs mappings for Android 2.2 framework using dynamic API fuzzing.
In contrast, PScout \cite{PScout12} employs a version-independent static analysis method to extract permission specifications from multiple versions of Android.

In this paper, we adopt API-to-permission mappings from Stowaway instead of PScout for two reasons.
First, although PScout provides a significant number of undocumented APIs, we only select VSinks from the range of documented APIs (see \S \ref{sec:problem}).
In terms of documented APIs, PScout does not provide more permission mappings than Stowaway.
Second, we find Stowaway produces a more accurate mapping than PScout, in terms of fewer false positives.
This might due to the fact that nearly every mapping found by Stowaway is verified by dynamic execution.
The accuracy of permission mapping is important for our VSink selection, because any incorrect VSink will directly introduce false positives to our ECV detection.
Therefore, currently we only adopt the mappings from Stowaway.
In the future, we could also take advantage of the version-independent feature of PScout.


We obtain a total of 456 validated documented APIs from Stowaway.
Assessing each API semantic is practically infeasible for two reasons:
(1) it would take a quite large manual workload,
and (2) two much manual analysis without a principle may incur some inaccuracy (e.g., assign wrong VSink tags).
Therefore, we propose to combine multiple metrics for systematic selection and classification.
These metrics include permission semantics and levels, API names, parameter and return-value types.
Among all metrics, permission semantic is the major metric for our system.
In fact, we only extract 56 kinds of permissions from those 456 documented APIs.
Therefore, we can divide them into 56 clusters (if an API needs multiple permissions, we choose its first marked permission), because APIs marked with the same permission are likely to share similar VSink nature.
For instance, combining with quick scan of API names, we find all 6 APIs under \texttt{SEND\_SMS} permission can be directly categorized into \DIRECTno.

\begin{figure}[h!]
  \centering
  \includegraphics[width=0.5\textwidth]{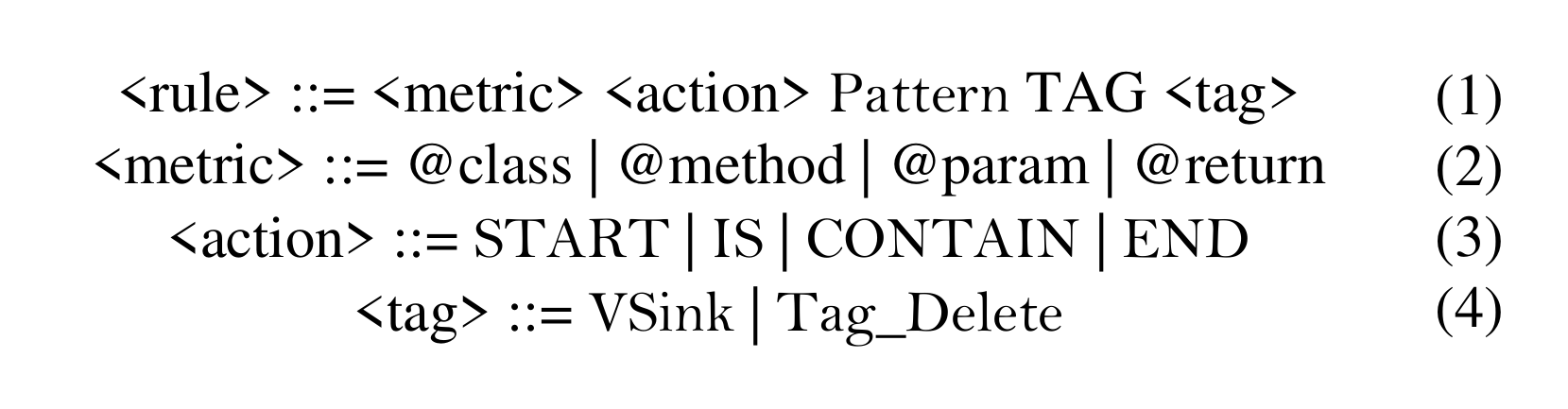}
  \vspace{-6ex}
  \caption{The rule syntax for VSink selection and classification.}
  \label{fig:rule}
\end{figure}

To facilitate multiple metrics based selection and classification, we further design a simple, but practical, rule syntax, as illustrated in \figurename \ref{fig:rule}.
Each rule is a five-element tuple, which describes what tag a particular API would be assigned, when one metric of this API satisfies a special string pattern.
We define four kinds of pattern-matching actions, such as ``START'' for ``start with'' action.
Finally, the tag in the syntax mainly includes four VSink categories.
Besides them, we also define a special tag named ``Tag\_Delete'', which can be used when we want to exclude an API.
Based on this syntax, we define four general rules and a set of permission-specific rules for extracting categorized VSinks from permission mappings.
We then write a rule interpreter to automatically select and classify VSinks according to the defined rules.

%


Besides the privileged APIs, another channel to select VSinks is the APIs that would make security impact on internal status of apps.
Three general kinds of such APIs have been considered in our work.
The first kind is the file operation APIs, such as the \texttt{write} and \texttt{append} methods from the \texttt{java.io.Writer} class.
Another kinds of APIs is the database-related APIs, such as those from \texttt{SQLiteDatabase} and \texttt{ContentResolver} classes.
The third one is the logcat APIs, i.e., logging APIs from the \texttt{android.util.Log} class.
We then design three additional rules to automatically join them into our VSink set.
Moreover, APIs from the \texttt{AndroidHttpClient} class (missed by both Stowaway and PScout) are similarly complemented into our final VSink results.


\subsection{Forward and Backward Analysis}
\label{sec:core}


\subsubsection{Forward Reachability Analysis}
\label{sec:forward}



We employ an iterative intra-procedural algorithm with flow sensitivity to perform reachability analysis (summarized in Algorithm \ref{alg1}). Note that since reachability analysis only relies on control flow, context-insensitive analysis is enough in our forward module. The algorithm first constructs a control flow graph (CFG) for each method, and then traverses every statement in a particular order according to the search strategy (e.g., DFS).
For each call site, we determine whether it is a VSink or can be resolved into a method defined by the app, respectively.
In particular, this call site resolving procedure is performed on demand along the forward analysis.
We maintain two lists for caching reached VSinks and resolved methods. After the lists stop growing, we obtain all reachable VSinks and then invoke specific modules to further process them. Finally, the algorithm will handle each resolved method in the same way.

\begin{algorithm}[t!]
\caption{Forward Reachability Analysis}
\label{alg1}
\begin{algorithmic}[1]


\Require $entryPts$ = [entry point], $vSinks$ = [VSink]
\ForAll {$ep \in$ $entryPts$}
    \State IterativeForward($ep$, $initchain$)
\EndFor
\State

\Procedure {IterativeForward}{$method$, $chain$}
\State $reachedS$ = [], $nextM$ = [], $chain$.add($method$)
\State

\State $cfg \leftarrow$ BuildCFG($method$)
    \ForAll {$cs \in$ CallSites(DFS($cfg$))}
        \If {$cs \in$ $vSinks$}
            \State $reachedS$.add($cs$)
        \Else
            \State $rc \leftarrow$ ResolveCall($cs$)
            \If {$rc \in$ [app defined method]}
                \State $nextM$.add($rc$)
            \EndIf
        \EndIf
    \EndFor
\State

\ForAll {$r \in reachedS$}
    \State $category \leftarrow$ JudgeCategory($r$, $vSinks$)
    \State InvokeSpecificModule($category$, $r$, $chain$)
\EndFor

\ForAll {$m \in nextM$}
    \State IterativeForward($m$, ShadowCopy($chain$))
\EndFor

\EndProcedure

\end{algorithmic}
\end{algorithm}

We generate call chains to facilitate inter-procedural backtrack analysis.
A call chain is a path of call graph, from an entry caller to an ending callee.
Generating individual call chains, instead of a whole call graph, could ease the path track procedure of subsequent modules and allow them concentrate on the design of flow analysis.
In contrast, traversal between two nodes on call graph might involve several parallel paths.
Moreover, an additional dataflow fact \textit{join} operation needs to be considered in the dataflow analysis using non-linear graph traversal.


Our call chain captures not only caller and callee methods, but also precise calling context information by recording call-site heap locations and their call strings (both are not shown in Algorithm \ref{alg1}, for simplicity).
On one hand, the heap location context information is essential for backward modules to jump back to each original call site.
Otherwise, ambiguity might arise when a caller method contains multiple similar call sites targeting the same callee.
On the other hand, we leverage call string to avoid re-analysis of callee method with the same dataflow value context.
More specifically, with the help of SSA IR form\footnote{SSA represents static single assignment, while IR is short for intermediate representation.}, we can distinguish call sites with different entry dataflow values through simply investigating their call strings.

\subsubsection{Backward Dataflow Analysis}
\label{sec:backward}

Except for \DIRECTno, the other three kinds of VSinks require further backward dataflow analysis.
Although each backward analysis is independent for a dedicated task, they share a common backward dataflow analysis method.
We thus first design this common backward method and then apply it to the three dedicated modules.

Our backward method relies on call chain (generated by forward module) to achieve inter-procedural context-sensitive backtrack analysis (shown in Algorithm \ref{alg2}).
The core of this algorithm is an iterative backward analysis procedure.
This iterative procedure starts with extracting SSA-form method body from call chain according to current chain index.
Then, it initializes a intra-method taint set and joins the incoming tainted variables.
After this, we need to locate the starting call site according to the provided calling context.
If calling context is the null (this can happen when we backtrack between two originally disconnected callback functions), we directly take the last call site as our starting point.
Similarly, we join the new tainted variables obtained from starting call site into the taint set.
We then loop all call sites before the starting point.
For each tainted call site, we further determine whether we need to propagate the taint into new variables.
Moreover, different dedicated modules may choose to mark result for tainted \SOURCE or constants at this time.
After the loop, some modules would further inspect whether there are tainted inputs.
Finally, the algorithm will judge whether it needs further backtracking, according to the current index number and variables in taint set.
For example, if we already backtrack to the first method in chain, or no parameters and fields tainted, the algorithm will terminate.

\algnewcommand{\LineComment}[1]{\(\triangleright\) #1}

\begin{algorithm}[t!]
\caption{Backward Dataflow Analysis}
\label{alg2}
\begin{algorithmic}[1]

\Require $cy$: category, $r$: a reached VSink call, $chain$, $vSrc$
\State $ii$ $\leftarrow$ $chain$.size()-1, $itv$ $\leftarrow$ GetTVarsFromCallSite($r$)
\State IterativeBackward($ii$, $itv$, $r$)
\Statex
\Statex

\LineComment $i$: index, $tv$: tainted variables, $cc$: calling context
\Procedure {IterativeBackward}{$i$, $tv$, $cc$}
\State $sb$ $\leftarrow$ $chain$.get($i$).GetSSABody()
\State $ts$ $\leftarrow$ InitTaintSet(), $ts$.join(GetTaintedVars($tv$))
\State $cs\_start$ $\leftarrow$ GetStartCallSite($cc$)
\State $ts$.join(GetTVarsFromCallSite($cs\_start$))

\State

\State $cs\_old$ $\leftarrow$ $cs\_start$
\While{$sb$.hasPreviousCallSite($cs\_old$)}
    \State $cs\_new$ $\leftarrow$ $sb$.getPreviousCallSite($cs\_old$)
    \If {isThisCallSiteTainted($cs\_new$, $ts$)}
        \State $ts$.join(Determine-PropagateTaint($cs\_new$, $ts$))
        \State MarkResult\_VSrc($cy$, $cs\_new$, $vSrc$)
        \State MarkResult\_Constant($cy$, $cs\_new$)
    \EndIf
    \State $cs\_old$ $\leftarrow$ $cs\_new$
\EndWhile
\State MarkResult\_Input($cy$, $ts$, $chain$)

\State
\If {isContinueIterative($i$, $ts$, $cy$, $chain$)}
    \State $lcc$ $\leftarrow$ $chain$.get($i$).GetLastCallContext()
    \State IterativeBackward($i$-1, TransformTVars($ts$), $lcc$)
\EndIf

\EndProcedure

\end{algorithmic}
\end{algorithm}

Generating proper taint objectives is important for our backward analysis.
First, we need to obtain appropriate initial taints from reached VSink calls.
Since our current VSinks have no fine-grained information to indicate which parameter is critical, we then take a conservative approach that taints all encountered parameters to avoid any false negative.
Moreover, some VSink APIs have no parameters involved (e.g., previously mentioned \texttt{HttpURLConnection.connect()}), we thus taint their caller object.
Second, we need to maintain a mapping for tainted parameters during the procedure of method switching, so that we can obtain the correct variable format under different SSA method bodies.

We further design three dedicated modules for tackling ECV detection problems under \DIRECTByno, \SENDno, and \PUBLICno.
With the help of proposed backward method, these modules are relatively easy to design.
Specifically, for \DIRECTByno, we mark the result from tainted constants and inputs.
Because sometimes static analysis cannot obtain accurate parameter values, e.g., when they rely on dynamic execution outputs.
Therefore, we adopt every tainted constants into \DIRECTBy result, to mimic the ideal values.
While for \SEND and \PUBLICno, our main task is to backtrack whether there are tainted inputs and \SOURCEno, respectively
.

\subsubsection{Analysis Enhancements}
\label{sec:enhance}

We also design three kinds of enhancements to the basic forward and backward analysis in \toolname.
In general, these enhancements can help \toolname reduce false positives, avoid unnecessary analysis overhead, and output more expressive result logs.

The first enhancement is to validate system-only broadcast checking in the flow analysis.
Broadcasts are system-wide events (e.g., battery is low), which will be delivered to registered Broadcast Receivers when the corresponding events occur.
For example, an Broadcast Receiver with the name \texttt{com.example.BootReceiver} (see \figurename \ref{fig:bootReceiver} in the Appendix) registers the \texttt{BOOT\_COMPLETED} broadcast, which will then trigger \texttt{BootReceiver} when the system has booted.
Note that although third-party apps can also declare their own broadcasts, many broadcasts are only sent by the operating system and prohibited by non-system senders \cite{Stowaway11}.
Therefore, attackers cannot inject a fake broadcast Intent with \texttt{BOOT\_COMPLETED} as the action name into \texttt{BootReceiver}.
However, it is still insufficient to prevent external crafted inputs, since attackers can directly trigger \texttt{BootReceiver} by set the explicit Intent target name.
A safe and common way to mitigate this issue is to explicitly check the action name of system broadcasts in the code, as suggested in \cite{ComDroid11} and \cite{ICCEpicc13}.
A typical code pattern for such system-only broadcast checking is illustrated in \figurename \ref{fig:checkBroadcast} (see Appendix).

Since broadcast checking can efficiently protect exposed Broadcast Receivers, we thus design a validation capability for system-only broadcast checking in \toolname's flow analysis, as an enhancement to the forward analysis in \S \ref{sec:forward}.
Besides helpful to reduce false positives, this validation can also improve the system performance, because \toolname can avoid the subsequent analysis once it identifies checking.
Our validation is targeted at the code pattern in \figurename \ref{fig:checkBroadcast} (i.e., the If-Else checking using \texttt{equals} API in Java's \texttt{String} class), since it is the most straightforward way to perform broadcast checking.
We also believe it is easy for \toolname to cover other kinds of checking patterns, once their domain knowledge is provided.

To facilitate accurate validation, a critical job is to collect sufficient system-only broadcasts.
Note that our goal is to identify enough broadcasts that could cover most app cases, and proposing a new way to dig out a complete set is out of the paper scope.
There are two existing related resources we can leverage.
First, the \texttt{AndroidManifest.xml} file in each version of Android source code defines a list of system broadcasts with the tag name \texttt{protected-broadcast}.
We thus collect 133 such broadcasts from the recent Android 4.3 platform.
Second, Stowaway \cite{Stowaway11} uncovers 62 system broadcasts by dynamic testing, including those dynamically declared in the code.
We then merge these two broadcast sets into a new one, and finally obtain in total of 143 unique system-only broadcasts for \toolname.

Second, we avoid backtracking some uncritical parameters to reduce overhead.
A typical example is the first \texttt{tag} parameter of logcat APIs, such as \texttt{Log.v(String tag, String msg)}.
Developers usually assign insensitive values (e.g., string constants) into this parameter, thus there is no need to analyze it.
Otherwise, the backward module may waste tracing several methods before arriving its initialization method.

Third, we output input-related variable values for more expressive result logs.
Our backward analysis could taint a dependence between the input and parameter, like CHEX \cite{CHEX12}.
However, such coarse-grained dependence without detailed propagation knowledge sometimes is not enough.
To this end, we choose to output some input-related variable values into our result logs, such as the log like \texttt{Input:r4 = r2.getStringExtra("referrer")}, where r2 is the original Input or Intent object.
In this way, we can obtain more expressive and meaningful result logs.

\subsection{Semi-auto Guided Analysis}
\label{sec:guided}

\begin{figure*}[ht!]
  \centering
  \includegraphics[width=0.86\textwidth]{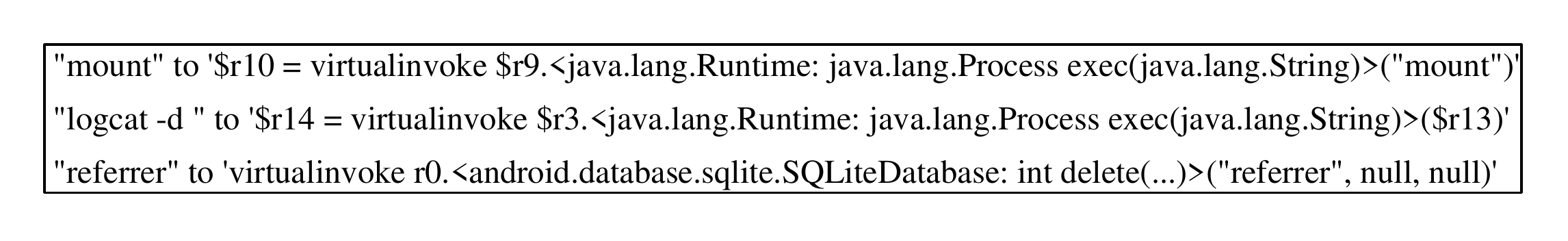}
  \vspace{-4ex}
  \caption{Logs showing example false positives in \DIRECTBy category.}
  \label{fig:guidedexample}
\end{figure*}

We need to further filter some false positives in the initial set of potential ECVs identified by previous analysis.
These false positives are mainly from two VSink categories: \DIRECTBy and \SEND.
The reason is that VSinks in these two categories generally rely on parameters to exhibit their specific behaviors.
Therefore, analyzing these VSinks is usually sink-specific and parameter-specific, meaning that we have to combine detailed sinks and their parameter values for detection.
However, it is nearly impossible for \toolname to automatically handle it, because too much domain knowledge is needed.

\begin{figure}[ht!]
  \centering
  \vspace{-3ex}
  \includegraphics[width=0.38\textwidth]{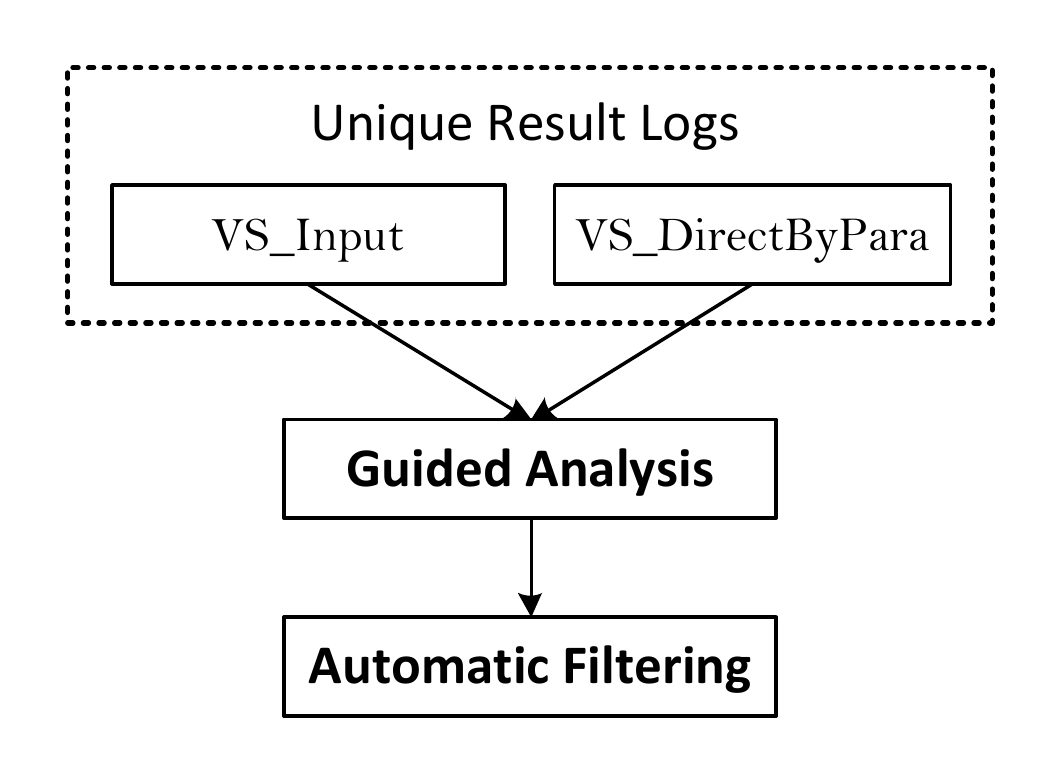}
  \vspace{-4ex}
  \caption{Basic process of semi-auto guided analysis.}
  \label{fig:guided}
\end{figure}

To address this issue, we propose semi-auto guided analysis to quickly filter false positives.
Its basic process is illustrated in \figurename \ref{fig:guided}.
In particular, we take advantage of the result logs outputted by \toolname.
Figure \ref{fig:guidedexample} shows the logs of example false positives in the \DIRECTBy category.
Specifically, we first collect all unique results logs from \DIRECTBy and \SEND.
Then, we conduct manual analysis to find all false positive logs and their patterns.
This step largely relies on expert knowledge.
Little effort is actually required because many logs are similar.
Finally, we apply our extracted false positive log pattern to all related apps and take scripts to automatically filter those matched cases.

\section{\toolname Implementation}
\label{sec:implement}

\begin{figure}[ht!]
  \centering
  \vspace{-4ex}
  \includegraphics[width=0.5\textwidth]{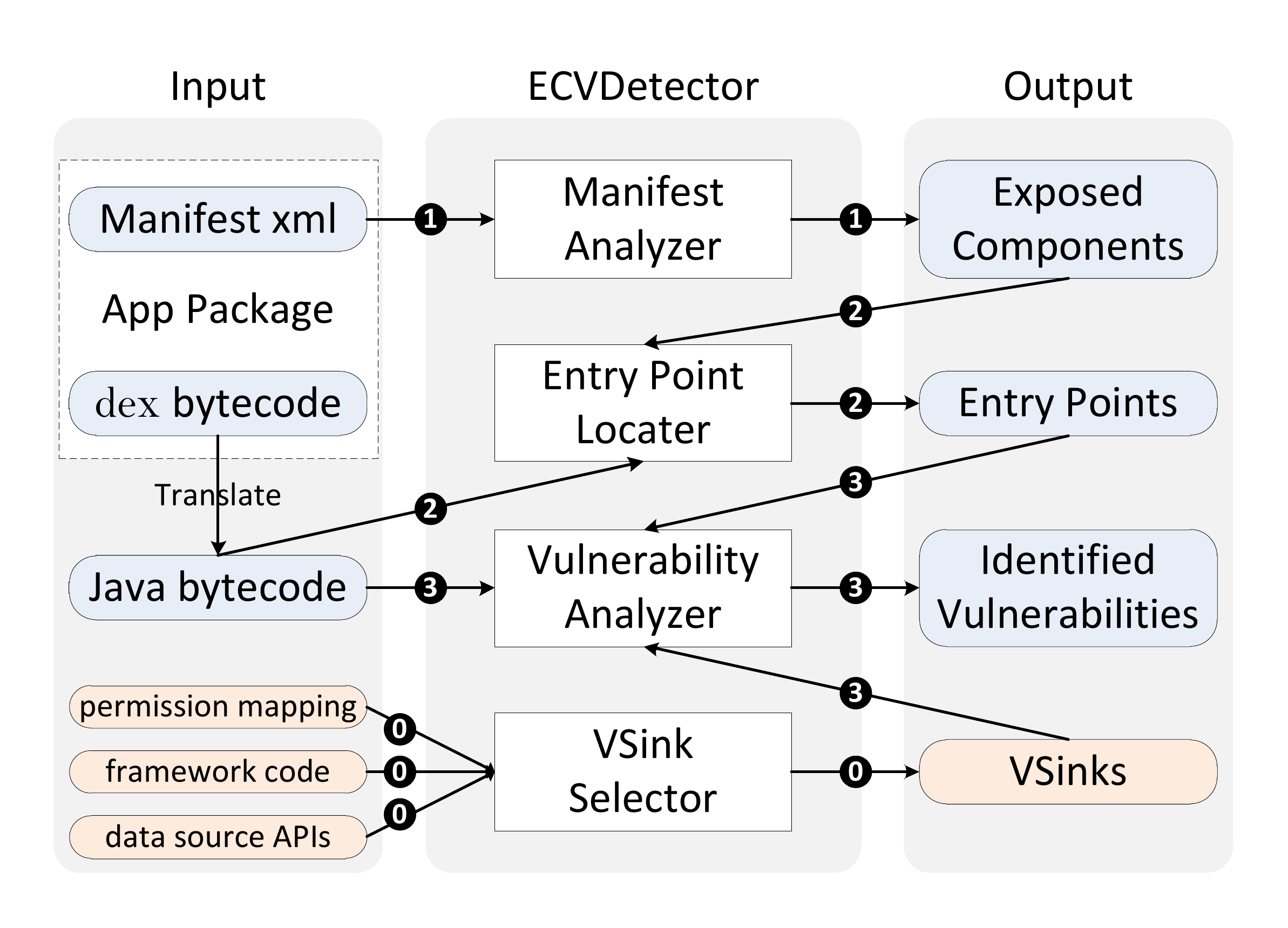}
  \vspace{-6ex}
  \caption{\toolname architecture. VSink Selector and Vulnerability Analyzer are two major components to implement our sink-driven approach.}
  \vspace{-4ex}
  \label{fig:arch}
\end{figure}

We have implemented \toolname with 4,565 lines of Java code, Python scripts and shell scripts.
As shown in \figurename \ref{fig:arch}, \toolname consists of four components.
Specifically, Manifest Analyzer is implemented only in Python scripts, while the other three components (i.e., Entry Point Locator, Vulnerability Analyzer and VSink Selector) are based on the Soot framework \cite{Soot11}.
\toolname contains four execution steps: one preparation step (i.e., step 0 at the bottom of \figurename \ref{fig:arch}) and three analysis steps.
The preparation step is executed only once for generating VSinks.
Then, \toolname analyzes each Android app for ECVs in three consecutive analysis steps.

\textbf{VSink Selector.}
By running VSink Selector with the inputs of Android framework code and Stowaway permission mappings, we obtain a total of 372 categorized VSinks.
Among them, 137 APIs are \DIRECT, 23 \DIRECTBy, 167 \SEND, and 45 \PUBLIC.
Moreover, to facilitate detecting ECVs involving sensitive sources, we also select 183 \SOURCE based on the results of Stowaway and SuSi \cite{SuSi13}.
As regards to the data sinks in SuSi, we find most of them cannot serve our vulnerability-specific purpose, due to their selection motivation for privacy leak detection.


To adopt the Stowaway mapping into our VSinks, however, faces a technical challenge that method signatures in Stowaway mapping are incomplete.
In fact, these incomplete signatures are only sub-signatures without return-value types, thus would incur inaccuracy for Vulnerability Analyzer.
To this end, we design several estimation rules to restore complete method signatures by analyzing corresponding Android framework code.
Moreover, since our analysis program requires right method sub-signatures to facilitate signature restoring, we unexpectedly identify several kinds of inaccuracy issues existing in Stowaway results.
These issues exist might because Stowaway also employs some manual efforts to produce the mapping, thus introduce some inaccuracies.
We then ran our analysis program several times to inspect the error messages (in each run), which prevents us to successfully restore all signatures.
For identified issues in each run, we manually fix them by batch replacing.

\textbf{Manifest Analyzer.}
This component consists of two parts: an xml parser to extract all essential information inside each manifest file, and a script to identify exposed components by analyzing extracted information.
We determine exposed components according to the rule in \S \ref{sec:exposed}.

\textbf{Entry Point Locator.}
The main task of Entry Point Locator is to locate all entry points, which would serve as the starting points for Vulnerability Analyzer.
Basically, entry points are fixed callback interfaces defined by Android programming paradigm, such as onCreate and onStart.
Particularly, onCreate is an initialization point which will be called when a component is started up.
Moreover, several kinds of entry points are special, in terms of they can accept external attack inputs.
While other points either take zero parameter (e.g., onResume and onStop), or only contain inputs cannot be manipulated by attacker (e.g., \texttt{Bundle} argument in Activity's onCreate entry point).
Since \SEND ECV detection is related to those special entry points, we identify them in Android framework and list in Table \ref{table:callback}.
Three kinds of components contain such entry points we concerned, while Activity needs to actively call \texttt{getIntent} fucntion to receive external inputs.

\begin{table}[ht!]
\centering
\small
\vspace{-3ex}
\caption{Our aimed entry point functions.}
\label{table:callback}
\renewcommand{\arraystretch}{1.0}	
\scalebox{0.95}{
\begin{tabular}{ c || l }
\toprule							
Component & Callback functions that accept attack inputs
\tabularnewline
\midrule
\multirow{5}{*}{Service} & onBind(\textbf{Intent} intent) \tabularnewline
		& onStart(\textbf{Intent} intent, int startId) \tabularnewline
	 	& onStartCommand(\textbf{Intent} intent, int flags, int Id) \tabularnewline
		& onHandleIntent(\textbf{Intent} intent) \tabularnewline
		& handleMessage(\textbf{Message} msg) \tabularnewline
\hline
Receiver& onReceive(Context context, \textbf{Intent} intent) \tabularnewline
\hline
\multirow{5}{*}{Provider} & query(\textbf{Uri}, String[], String, String[], String) \tabularnewline
		& insert(\textbf{Uri}, ContentValues) \tabularnewline
		& update(\textbf{Uri}, ContentValues, String, String[]) \tabularnewline
		& delete(\textbf{Uri}, String, String[]) \tabularnewline
		& openFile (\textbf{Uri}, String) \tabularnewline
\bottomrule
\end{tabular}
}
\end{table}

Another task is to model lifecycle for identified entry points.
This is necessary because the entry points will not call each other due to their callback nature, thus cause disconnected static flows.
Moreover, there are some initialization functions even before the onCreate interface, such as $<$clinit$>$ and $<$init$>$.
We also need to consider them in modeling lifecycle, so that backward module could backtrack to the initial definition.
In the current \toolname, we model the lifecycle by defining several continuous phases.
Each phase may contain a manually connected flow or several asynchronous entry points.
For example, we define an ``initial'' phase to connect the initial flow (i.e., $<$clinit$>$ $\rightarrow$ $<$init$>$ $\rightarrow$ onCreate),
and a ``main'' phase to cover all asynchronous special entry points in Table \ref{table:callback}.


\textbf{Vulnerability Analyzer.}
This component mainly performs the sink-driven forward and backward analysis in \S \ref{sec:core}.
In the process of implementing Vulnerability Analyzer, we come across two major technical issues due to:
(1) object-oriented (OO) language used by Android development,
and (2) Android's event-driven nature.

The first issue arises from the inheritance nature of OO language during the call site resolving in the forward module.
To obtain the target method of a call site, it is essential to resolve the type of target class object.
However, due to the inheritance, it is hard to statically determine what concrete class an object would represent.
More specifically, the inheritance nature allows developer to use superclass or interface type to represent a subclass object, which causes the ambiguity.
We tackle this problem to a great extent by leveraging typed Soot IR \cite{Soot11}, which is calculated by a fast type inference algorithm \cite{OOPSLA08}.
Besides the object type resolving, OO language's inheritance nature also makes locating the right method definition complex, even if we have obtained the exact object type.
For example, \texttt{toString} method invoked by \texttt{android.graphics.Bitmap} object is actually not defined by \texttt{Bitmap} class itself.
Indeed, we have to track back to \texttt{java.lang.Object} class to obtain the method definition of \texttt{toString}.
A straightforward way for mitigating this problem is to maintain a comprehensive class hierarchy.
In our \toolname prototype, we also take advantage of the automatic method resolution provided by Soot according to Java Virtual Machine Specification \cite{JVM7}.

The second issue is the static flow discontinuity problem caused by Android's event-driven nature \cite{Woodpecker12}.
A typical disconnected example is the flow between \texttt{start()} and \texttt{run()} method in \texttt{java.lang.Thread} class.
In fact, Android OS connects these two methods dynamically through the underlying thread scheduler.
However, a static tool, without specific knowledge, cannot directly predict this kind of dynamic connecting behavior.
Nevertheless, it is necessary for Android static analysis tools to tackle this problem in a satisfactory way.
Otherwise, some false negatives would arise.
In \toolnameno, we model a number of dynamic flow connecting behaviors as pre-defined knowledge to build continuous call chains.
The main modeled flow connecting behaviors are summarized in Table \ref{table:model}.
According to this table, \toolname is capable to connect the disconnected flows occurred in thread scheduling, timers, location updates, and so on.

\begin{table}[htb]
\centering
\vspace{-3ex}
\caption{The main dynamic flow connecting behaviors modeled by \toolnameno.}
\label{table:model}
\renewcommand{\arraystretch}{1.0}	
\begin{tabular}{ c || l }
\toprule							
Class name & Modeled Flows
\tabularnewline
\midrule
Thread & start $\rightarrow$ run \tabularnewline
AsyncTask & execute $\rightarrow$ doInBackground \tabularnewline
Handler & sendMessage $\rightarrow$ handleMessage \tabularnewline
Timer & schedule $\rightarrow$ run \tabularnewline
LocationManager & requestLocationUpdates \tabularnewline
                & $\rightarrow$ \emph{LocationListener} \tabularnewline
TelephonyManager & listen $\rightarrow$ \emph{PhoneStateListener} \tabularnewline
\bottomrule
\end{tabular}
\end{table}

\section{Evaluation}
\label{sec:evaluate}

%

To evaluate the efficacy and performance of \toolname, we carried out an evaluation with top 1K Android apps from Google Play.
The reason we evaluated these popular apps is that the impact of ECVs relies on the popularity of vulnerable apps.
In other words, vulnerable apps with few installs would not cause big security impact, because it is less likely to let those limited users also install the attack apps.
Specifically, these top apps were selected according to their user review numbers, and most of them were crawled recently (between June and July 2013).
Therefore, our app dataset could represent the recent versions of top Google Play apps that users may install in their phones.
Moreover, our selected apps were fully unique in terms of the package names, which made our dataset more distributed.


In this section, we first discuss how we conducted the experiment with this dataset, including some findings and knowledge we obtained during the procedure of experiment.
Then we report our identified ECVs and conduct case studies of some representative vulnerable apps.
Finally, we depict the result of performance evaluation for \toolname.

\subsection{Experiment and Findings}
\label{sec:experiment}

One essential step before the \toolname experiment is to extract the AndroidManifest.xml for each app.
As mentioned in \S \ref{sec:implement}, we took apktool for unpacking the app and obtaining the manifest file.
However, we found not all apps could be correctly handled by apktool.
Indeed, six apps of our dataset fail and crash apktool with several Java exceptions, maybe due to the potential bugs in apktool or some apps try to protect itself from the decompiling of apktool \cite{SlowDown13}.
This finding also gives a kind hint to all previous work based on apktool, such as \cite{Elish13} and \cite{RiskRanker12}.

We then leveraged the Manifest Analyzer component of \toolname to discover exposed components from the rest of 994 apps.
Among them, we successfully parsed and analyzed 992 manifest files.
The two failure cases were due to the invalid encodings that cannot be handled by the Python XML DOM library.
More specifically, one failed case (the popular Titanium Backup app) contains the Chinese and Korean characters for some component names.
These two failed manifest files could be manually analyzed for obtaining exposed components, but in this paper we just ignored them for the automatic experiment.

\begin{figure}[ht!]
  \centering
  \vspace{-1ex}
  \includegraphics[width=0.45\textwidth]{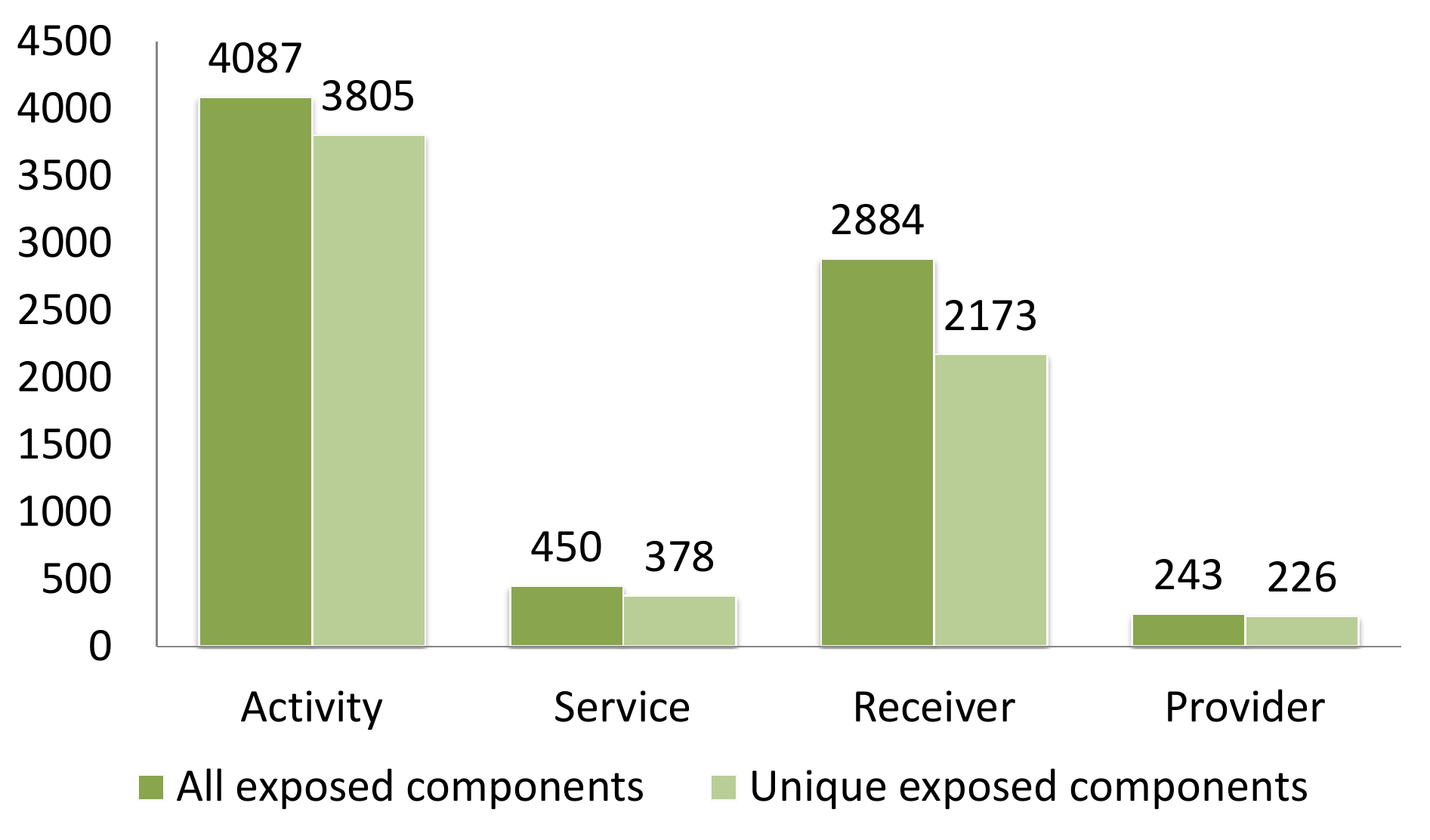}
  \vspace{-2ex}
  \caption{The amounts of all and unique exposed components across four component types.}
  \label{fig:exposed}
\end{figure}

In summary, we identified a total of 7,664 exposed components from the remaining 992 apps, and 6,582 of them were unique in terms of the component name.
The detailed amounts of exposed components classified by component types are illustrated in \figurename \ref{fig:exposed}.
One major finding based on this figure is that there are significantly more Activity and Broadcast Receiver exposed components than the other two component types, around ten times.
The reason behind this can be explained by these two facts.
First, many apps need exposed Activities to finish the user intention of app switching, such as launching from the launcher app.
Second, in order to receive system broadcasts, the corresponding Receivers have to expose themselves to the Android framework.

Our experiment for ECV detection focused on the exposed Services and Broadcast Receivers.
More specifically, our target was 2,551 unique exposed components, including 378 Services and 2,173 Receivers (see \figurename \ref{fig:exposed}).
The reason for skipping Activities is mainly because exploiting Activity vulnerabilities usually requires user's intention (e.g., the Adobe Activity vulnerability shown in \cite{DroidChecker12}), and the complicated attacks against Browser Activities \cite{CrossOrigin13} are out of the scope of our paper (see \S \ref{sec:problem}).
As for Content Providers, although \toolname is capable of statically detecting some potential Provider vulnerabilities, heavy manual efforts are still required even with the dedicated ContentScope \cite{ContentScope13} tool.

Next, we ran \toolname against those 2,551 exposed components.
The experiment was performed on a single Dell PowerEdge storage server, equipped with four 2.40GHz CPUs and 12GB of RAM.
To optimize the throughout, we set the timeout for processing each component to two minutes.
This threshold was quite reasonable, because we found most of components could be analyzed within 10 seconds (see \S \ref{sec:performance} later).

During the experiment, we found it was challenging to prevent Soot from crashing when loading Java classes for analysis.
We spent many efforts to make \toolname successfully analyze all 2,551 components, and our knowledge listed as follows might be helpful other Soot-based Android tools \cite{FlowDroid13, ICCEpicc13, AppSealer14}.
The first knowledge is that we should provide as many underlying Android classes as possible to Soot.
These classes would help Soot resolve most of Android-specific types.
Specifically, we provided eight platforms of Android SDKs to Soot, from the old 1.6 to the recent 4.3.
Moreover, two platforms of hidden and system Android APIs were also provided to Soot, as well as one version of the Google APIs (e.g., Google Map APIs).
However, we still encountered many failed cases due to missing app-specific classes.
In order to solve this issue, we made use of the second knowledge:
ask Soot to only load classes on demand, and if errors still occur then set the Soot option \texttt{allow\_phantom\_refs}.
This option would let Soot create a phantom class for each missing class.

Another interesting finding is about system-only broadcasts, which are discussed in \S \ref{sec:enhance}.
Our finding about these broadcasts for the 2173 exposed Receivers is divided into three parts.
First, we identify 433 of them contained the checking for system-only broadcasts in their codes.
This result suggests our designed validation capability (\S \ref{sec:enhance}) could effectively alert the false positives due to system-only broadcast checking.
Second, total 40 broadcasts of our selected 133 system-only broadcasts are checked.
Third, we also count the numbers of checked broadcasts, and the top three result is listed in Table \ref{table:topbroadcast}.
The fourth most checked broadcast is \texttt{TIMEZONE\_CHANGED}, but with only 19 times.

\begin{table}[ht!]
\centering
\small
\vspace{-3ex}
\caption{Top 3 checked system-only broadcasts.}
\label{table:topbroadcast}
\renewcommand{\arraystretch}{1.0}	
\begin{threeparttable}
\begin{tabular}{ c || c }
\toprule							
The Checked System-only Broadcast & \# \tabularnewline
\midrule
android.intent.action.BOOT\_COMPLETED & 157 \tabularnewline
\midrule
android.net.conn.CONNECTIVITY\_CHANGE & 57 \tabularnewline
\midrule
android.intent.action.PACKAGE\_ADDED & 43 \tabularnewline
\bottomrule
\end{tabular}
\end{threeparttable}
\end{table}

\begin{table*}[ht!]
\centering
\vspace{-1ex}
\caption{Four categories of identified ECVs and their representative vulnerable apps.}
\label{table:confirmed}
\renewcommand{\arraystretch}{1.0}	
\scalebox{1}{
\begin{threeparttable}
\begin{tabular}{ c | c || c | l }
\toprule							
\multirow{2}{*}{ECV Category} & \# of & \multicolumn{2}{c}{Representative Apps} \tabularnewline
\cline{3-4}
& Apps & Package Name & Vulnerability Description \tabularnewline
\toprule							
\multirow{4}{*}{\DIRECT} & \multirow{4}{*}{6} & com.jb.gosms & Force it to send SMS to phone no. specified by input \tabularnewline
\cline{3-4}
& & com.gau.go.launcherex. & \multirow{2}{*}{Force it to enable or disable wifi and bluetooth} \tabularnewline
& & gowidget.switchwidget & \tabularnewline
\cline{3-4}
& & com.bwx.bequick & Force it to open the camera as flash light \tabularnewline
\midrule
\multirow{3}{*}{\DIRECTBy} & \multirow{3}{*}{5} & com.cleanmaster.mguard & Force it to silently uninstall arbitrary apps and etc. \tabularnewline
\cline{3-4}
& & mominis.Generic\_ & \multirow{2}{*}{Force it to delete its internal ``MESSAGE'' database table} \tabularnewline
& & Android.Ninja\_Chicken & \tabularnewline
\midrule
\multirow{6}{*}{\SEND} & \multirow{6}{*}{25} & com.zlango.zms & Force it to change the status of messages in SMS databases \tabularnewline
\cline{3-4}
 & & com.antivirus & Start the private core AVService with extras specified by input \tabularnewline
\cline{3-4}
& & com.doubleTwist. & \multirow{2}{*}{Start a private service with dangerous action named ``delete\_db''} \tabularnewline
& & androidPlayer & \tabularnewline
\cline{3-4}
& & com.linkedin.android & Launch a network request with the attributes specified by input \tabularnewline
\cline{3-4}
& & com.ebuddy.android & The beta update Receiver can be cheated to download fake apps \tabularnewline
\midrule
\multirow{5}{*}{\PUBLIC} & \multirow{5}{*}{13} & com.symantec. & \multirow{2}{*}{getLastKnownLocation() is outputted to the logcat} \tabularnewline
 &  & mobilesecurity & \tabularnewline
\cline{3-4}
& & air.com.bitrhymes.bingo & getDeviceId() along with post URL are outputted to the logcat \tabularnewline
\cline{3-4}
& & com.levelup.touiteur & getAllNetworkInfo() is outputted to the logcat \tabularnewline
\cline{3-4}
& & com.sec.spp.push & getConnectionInfo() is outputted to the logcat \tabularnewline
\bottomrule
\end{tabular}
\end{threeparttable}
}
\end{table*}

Through the automatic analysis for the 2,551 exposed components of 992 apps, we discovered the initial set of potential ECVs with 348 affected apps.
As previously shown in \figurename \ref{fig:workflow}, this result was analyzed only by the forward and backward analysis in \toolname.
We still needed to perform the semi-auto guided analysis for two ECV categories (\DIRECTBy and \SEND), as discussed in \S \ref{sec:guided}.
Specifically, there were 172 and 357 unique logs (need guided analysis) for \DIRECTBy and \SEND, respectively.
Nevertheless, most of them were similar to each other, only 6 kinds of \DIRECTBy APIs and 16 kinds of \SEND APIs were identified.
Therefore, we did not spend much effort (around one hour in total) in validating these logs.
\toolname then automatically filtered various sink-specific false positives according to our extracted log pattern.
Eventually, we identified a total of 103 potential vulnerable apps.
More specifically, there are 31 apps affected with potential \DIRECT ECVs, 25 for \DIRECTBy, 56 for \SEND, and 16 for \PUBLIC.


\subsection{Identified ECVs}
\label{sec:evaluate_vulns}

The 103 potential vulnerable apps (identified in the last section) were further manually verified for the real ECVs, due to the lack of ground truth.
The result of this manual verification mainly relies on the domain knowledge of security analysts and their defined boundary about bug and vulnerability, as discussed in \S \ref{sec:problem}.
In our verification process, we tried to follow a relatively more conservative principle than the closest related work CHEX \cite{CHEX12}.
A typical example is that the vulnerable case of ``Object embedded in input used to start Activity'' in CHEX would not be directly treated as a vulnerability in our principle.
Instead, we also require this action of starting Activity could make some security impact to the internal status of victim apps for becoming an ECV.
Some related preliminary knowledge and discussion could be also found in \S \ref{sec:problem}.

\noindent \textbf{Overall results.}
In the end, we tagged 49 apps as vulnerable from 103 candidate ones.
We confirmed them mainly by carefully auditing the decompiled codes and manifest files (with the help of result logs and call chains produced by \toolname), similar to the verification used by CHEX.
Note that for each vulnerable app, we only tagged it to one major ECV category, even if this app might be affected by several ECVs.
The main result is summarized in Table \ref{table:confirmed}, including the number of each category of identified ECVs and their corresponding representative vulnerable apps.
To the best of our knowledge, most of them are not disclosed previously, thus are the zero-day ECVs.
Since these vulnerable apps are from top 1K in Google Play, their vulnerabilities may incur real security consequences among many users.
We have reported several particularly serious cases to corresponding vendors, and received the acknowledgements from Go Dev team and LinkedIn.

The true positive rate of \toolname is not high (i.e., 47.6\%, 49/103), but we argue this is acceptable when consider the following two factors.
The first factor is the aforementioned conservative verification principle we used, which cause us to report less vulnerabilities.
Some \SEND APIs are the most affected by this principle, mainly the IPC APIs (e.g., \texttt{startActivity} and \texttt{startService}).
This problem could be migrated when further cross-component inspection is involved.
The second is due to the characteristics of some our selected sinks.
For instance, the \texttt{removeAccount} \DIRECT API is commonly used in a reasonable scenario---when an user logs out her account, the corresponding app then removes her local account cache from the phone---which should not be treated as an vulnerability.
However, it is hard for \toolname to recognize this normal situation of API usage.

\noindent \textbf{Case studies.}
We now conduct three cases studies to demonstrate \toolname's capability of detecting real serious ECVs under different categories. 
For each case, we further write an exploit in a zero-permission app to confirm the corresponding ECV could be effectively exploited.
Automatically generating exploits is an interesting topic for aiding static analysis, but is out of our paper scope and will be a future work.
Here we hide app version information in case that some users are still using the vulnerable versions of apps.

Go SMS Pro (\texttt{com.jb.gosms}) is the top 1 messaging app with over 60 million installs.
\toolname identified its exposed \texttt{CellValidateService} component leaked a security-critical flow path which arrived at the \texttt{sendTextMessage} sink API under the \DIRECT category. 
Moreover, its first parameter for assigning target phone number is completely controllable by external Intent.
Thus, a zero-permission attack app can force Go SMS Pro to send SMS to arbitrary phone number.
We also prepared a demo video (at \url{https://www.youtube.com/watch?v=CwtNCwAHSRs}) and reported it to Go Dev team on Sep 9 2013.
From their active response, we knew we were the first reporter on this issue.
This demonstrates the efficacy of \toolname, even when \texttt{sendTextMessage} is a commonly-used sink. 

Clean Master (\texttt{com.cleanmaster.mguard}) is a quite popular clean up app with over 200 million worldwide installs.
\toolname identified an external Intent can control its exposed \texttt{LocalService} to do privileged operations, such as clean up memory, restore apps, and silently uninstall arbitrary apps.
Take silently uninstalling apps as an example, the flow to a \DIRECTBy command execution API could be injected into attacker-supplied parameters (e.g., a victim app).
Clean Master then executes the ``pm uninstall'' command to do silent uninstallation (of course, root permission needs to be pre-granted).
This case demonstrates \toolname can uncover serious \DIRECTBy vulnerabilities.

Lango Messaging (\texttt{com.zlango.zms}) is another top messaging app with over one million installs.
\toolname identified an external Intent (at its \texttt{data} field) can flow into a \SEND API (called \texttt{ContentResolver.update}) via the exposed \texttt{ZmsSentReceiver} component.
We further found this exposed sensitive flow can enable a zero-permission app to maliciously change the status of some SMS messages.
For example, a draft SMS can be changed into the ``sent'' status and this changing will be reflected in the UI of all installed messaging apps.
Even worse, an incoming SMS can be set as an outgoing message, thus indirectly producing fake SMS messages.
This case shows the capability of \toolname to detect \SEND ECVs.

However, generating an effective exploit for Lango Messaging is surprisingly challenging, nearly taking us one-day effort.
The major difficulty is how to obtain an appropriate \texttt{Uri} data field from infinite candidates with the format of ``\texttt{content://}''.
The only hint is a domain knowledge required condition judgment (\texttt{MessageItem.getBoxId() != 2}).
Eventually, we figure out the \texttt{MessageItem} object (from \texttt{MessageItemManager.get(Uri)}) refers to a SMS message, and its status should be not as ``sent''.
Therefore, a target \texttt{Uri} can be ``\texttt{content://sms/inbox}||\texttt{draft/id}''.
The next step is to calculate a right message id.
Since the attack app has no \texttt{READ\_SMS} permission, it needs to launch brute-force attempts.
In summary, this case shows static detection tools (like \toolname) and manual analysis are still necessary. 
By contrast, automatic exploit generation (like fuzzing \cite{JarJarBinks12, IntentFuzzer14} or symbolic execution \cite{Diordna12}) is nearly impossible for handling this case, due to the lack of domain knowledge and setting up related system conditions (in this case, some incoming and draft SMS have to be prepared).

\PUBLIC ECVs are relatively not so serious, and require more app-specific conditions to trigger the vulnerable paths.
Therefore, we skip their case studies in this paper.

\subsection{Performance Evaluation}
\label{sec:performance}

\begin{figure}[t!]
  \centering
  \includegraphics[width=0.48\textwidth]{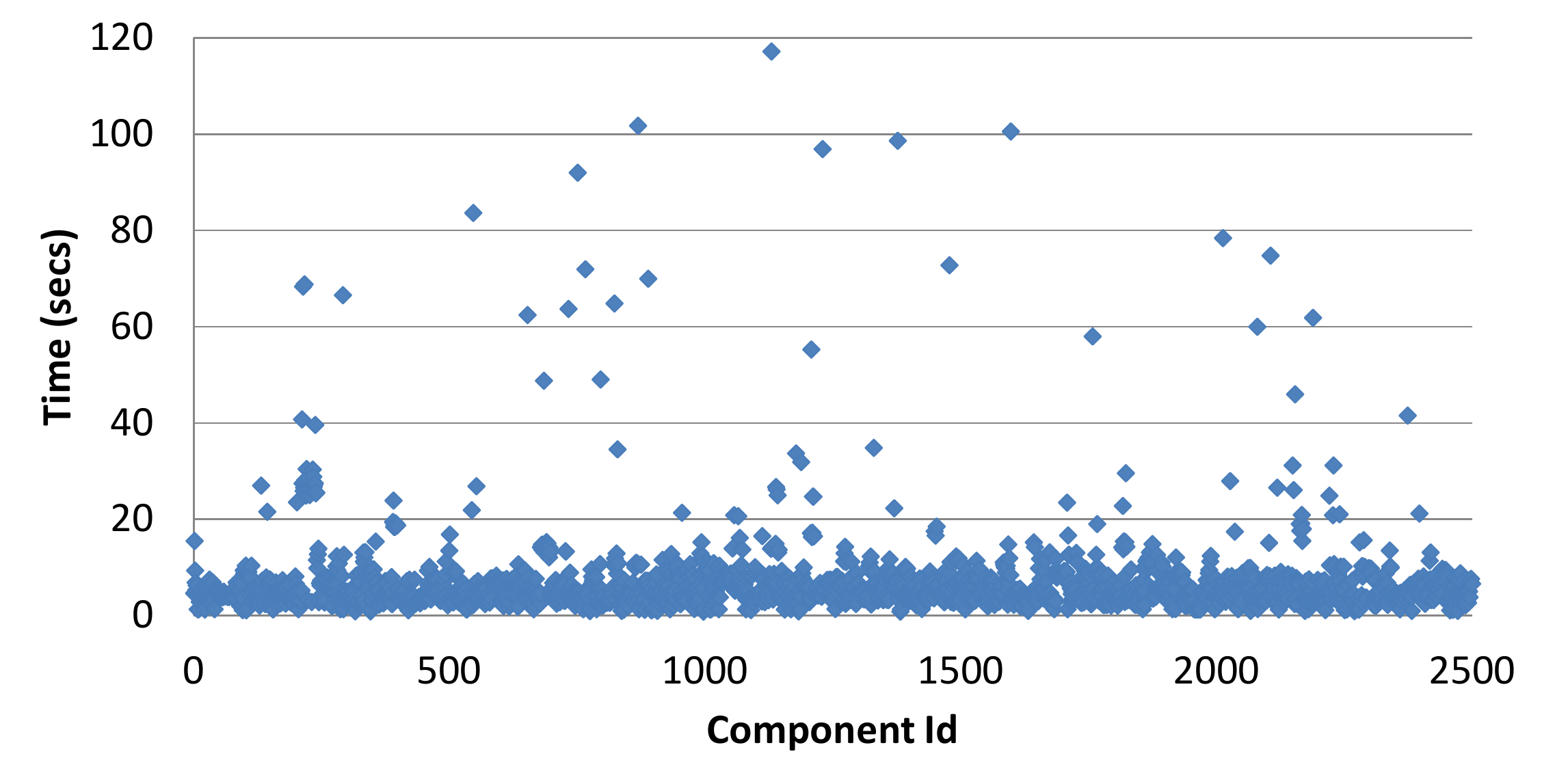}
  \caption{Detailed performance measurement.}
  \label{fig:performance}
\end{figure}

\toolname spent 6h 33m 35s to analyze the total 2,551 exposed components.
Therefore, the average processing time for each component was 9.257s.
This result suggests the threshold time of two minutes is quite reasonable.
Among the 2,551 components, we found only 35 of them arrived this timeout value.
The reason might be the complicated recursion codes that \toolname cannot recognize, although we have designed some mechanisms to help \toolname handle simple recursion.
We further give the detailed performance measurement for the rest of 2,516 components, as shown in \figurename \ref{fig:performance}.
It is easy to determine that most of components could be analyzed within 10 seconds.
Therefore, we conclude that the performance of \toolname is high.

\section{Discussion}
\label{sec:discuss}

The false positives shown in the last section are mainly due to the variable semantics of the \DIRECTBy and \SEND APIs.
It is hard to reduce these false positives even with dynamic analysis, because they also cannot determine which parameter values are critical and harmful. 
However, security analysts can extract the pattern of false positives from existing manual analysis, and feed them into the automatic filtering module of our guided analysis (\S \ref{sec:guided}).
In this way, some false positives could be further excluded automatically.

Similar to other works, there are also some false negatives in the current \toolname prototype.
However, it is hard to measure their quantity due to the lack of ground truth.
Thus, we discuss several possible causes for these false negatives as follows.
First, our modeling for connecting dynamic flows (see Table \ref{table:model}) is not intended to be complete, but we have tried our best to cover the common cases, similar to \cite{Woodpecker12, ContentScope13}.
Second, although we have proposed a systematic strategy to collect many sinks, it is impossible to cover a complete set.
Third, our current implementation of backtracking call chains is also not prefect in that we skip analyzing other callee methods in the same level of hierarchy for code simplicity and performance consideration. 
Fortunately, only a small number produce false negatives is due to this limitation.
Finally, the cross-component and cross-app ECVs \cite{CustomRom13} are not considered in our work, but we could leverage Epicc \cite{ICCEpicc13} to build a knowledge database and handle these cases.

\section{Related Work}
\label{sec:related}

%
%
\textbf{Exposed Component Vulnerability.}
This class of vulnerabilities has attracted much research effort recently, after Davi et al. \cite{ISC10_Privilege} presents the basic ECV model focusing on permission leaks.
Subsequently, several detection methods with different foci have been proposed.
One category of these methods includes ComDroid \cite{ComDroid11} and Epicc \cite{ICCEpicc13}, both of them try to identify several potential security flaws during the Intent-based communication.
As mentioned in \S \ref{sec:problem}, they aim at discovering all attack surfaces and give security warnings.
Therefore, they do not conduct sink-based analysis as ours.
As a result, they issue many potential ECV warnings, but it is hard to uncover true ECVs that have security impacts.

Another category of methods performs sink-based flow analysis to identify vulnerabilities more accurately.
Specifically, Woodpecker \cite{Woodpecker12} and DroidChecker \cite{DroidChecker12} leverage path reachability analysis to detect capability or permission leaks, which generally belong to our \DIRECT ECV category except for the non-ECV vulnerabilities on unauthorized Intent receipt.
SEFA \cite{CustomRom13} further extends this idea by adopting content leak detection \cite{ContentScope13} in exposed Content Providers.
CHEX \cite{CHEX12}, on the other hand, employs inter-procedural and context-sensitive dataflow techniques to locate suspicious ECV flows terminating at the data sinks (i.e., our \SEND and \PUBLIC APIs).

Similar to other sink-based analysis systems, \toolname also conducts path reachability and dataflow analysis.
However, unlike previous works, these two kinds of analysis techniques are organically combined together and driven by the classified categories of VSinks in our approach.
In particular, we take the backward, instead of previous forward, dataflow analysis for adapting more categories of sinks, such as the \DIRECTBy category.
These categorized VSinks are obtained by our systematic VSink selection strategy, which also assists \toolname to cover more ECVs that are not addressed by previous work.
Additionally, we further design semi-auto guided analysis and system-only broadcast checking capability to efficiently exclude some false positives.



Besides detection techniques, several defense methods are proposed to mitigate ECV issues.
In general, they are dynamic enforcing systems, therefore required to modify Android source code.
They propose to check IPC call chains \cite{IPCInspection11, Quire11}, or even other channels like sockets and files \cite{Taming12}.
On the other hand, Kantola et al. \cite{IntraComDroid12} tries to automatically reduce unnecessary exposed surfaces.
Moreover, mandatory access control is tailored to Android \cite{SEAndroid13, FlaskDroid13}, and it could block ECV attacks with appropriate policies.
Quite recently, AppSealer \cite{AppSealer14} aims to automatically generate patches for preventing attacks to exploit ECVs. 
All these works are complementary to our work, since they might be applied to prevent attackers from exploiting our discovered ECVs.

\textbf{Sink Selection in App Analysis.}
Many app analysis tools rely on sinks to perform their individual analysis.
However, most of their sinks are selected in a manual fashion \cite{TaintDroid10, AppFence11, PiOS11, Woodpecker12, ContentScope13}.
Some works make use of API-to-permission mappings \cite{Stowaway11, PScout12} to obtain their target sinks, such as \cite{AndroidLeaks12, AdRisk12, CHEX12}.
However, due to the lack of systematic strategy and flexible rules like our own, it is not easy for them to systematically extract appropriate sinks and filter useless ones from all candidate mappings.
Moreover, we craft centralized rules to adopt the privileged APIs not covered by existing mappings, as well as other non-privileged APIs like database APIs.
Another related work is SuSi \cite{SuSi13}, which employed machine learning techniques to select data sinks for privacy leak detection.
However, we cannot use SuSi to select our VSinks because of the variable API semantics and that we also consider non-data sinks.

A contribution of this paper is the semi-auto VSink classification.
SuSi and CHEX also define categories for their selected sinks, but their categories are not meant for capturing different analysis requirements like ours.
In fact, the categories in SuSi are only the API types (e.g., Network and File sinks).
Similarly, CHEX defines three sink tags for just differentiating different data sinks.
On the other hand, although both DroidChecker \cite{DroidChecker12} and AdRisk \cite{AdRisk12} separate their sinks into two categories for analysis, they are not fully aimed at detecting ECV detection, thus not sufficient for all analysis requirements in ECV detection.
Indeed, one of their categories is either for detecting unauthorized Intent receipt or for privacy leak detection.


\section{Conclusion}
\label{sec:conclude}


In this paper, we presented a new sink-driven approach to systematically tackle the ECV detection problem.
This approach includes a systematic strategy for VSink selection and classification, and a general detection method to identify potential ECVs in Android apps.
We implemented our sink-driven approach in a tool called \toolname.
We successfully identified a total of 49 vulnerable apps across all four ECV categories in the top 1K Android apps.
Future works include helping developers fix their vulnerable apps and deploying \toolname as a web-based detection service.

%
\bibliographystyle{abbrv}
\bibliography{asiaCCS}
%

\appendix

\section{System-only Broadcast Checking}

\begin{figure}[htb]
  \centering
  \vspace{-3ex}
  \includegraphics[width=0.45\textwidth]{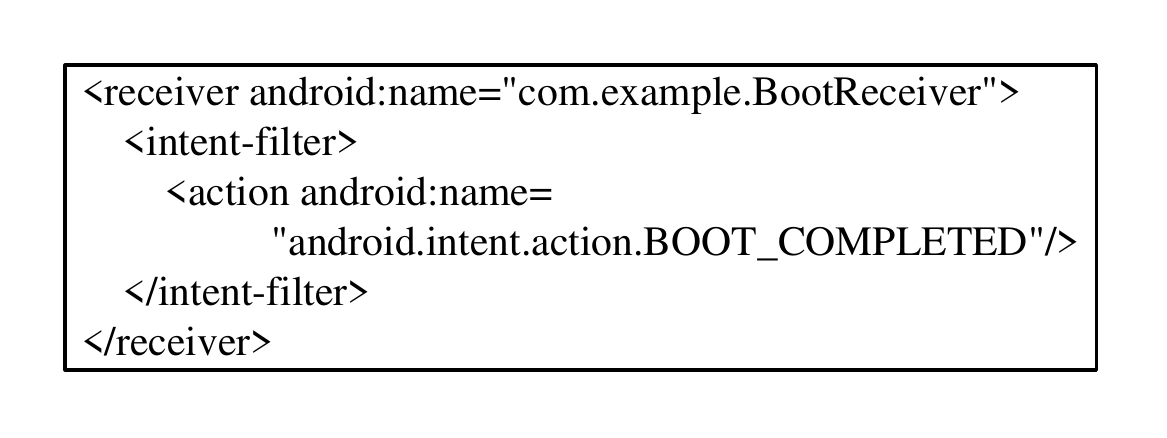}
  \vspace{-6ex}
  \caption{An example Broadcast Receiver registering the \texttt{BOOT\_COMPLETED} broadcast.}
  \label{fig:bootReceiver}
\end{figure}

\begin{figure}[htb]
  \centering
  \vspace{-3ex}
  \includegraphics[width=0.45\textwidth]{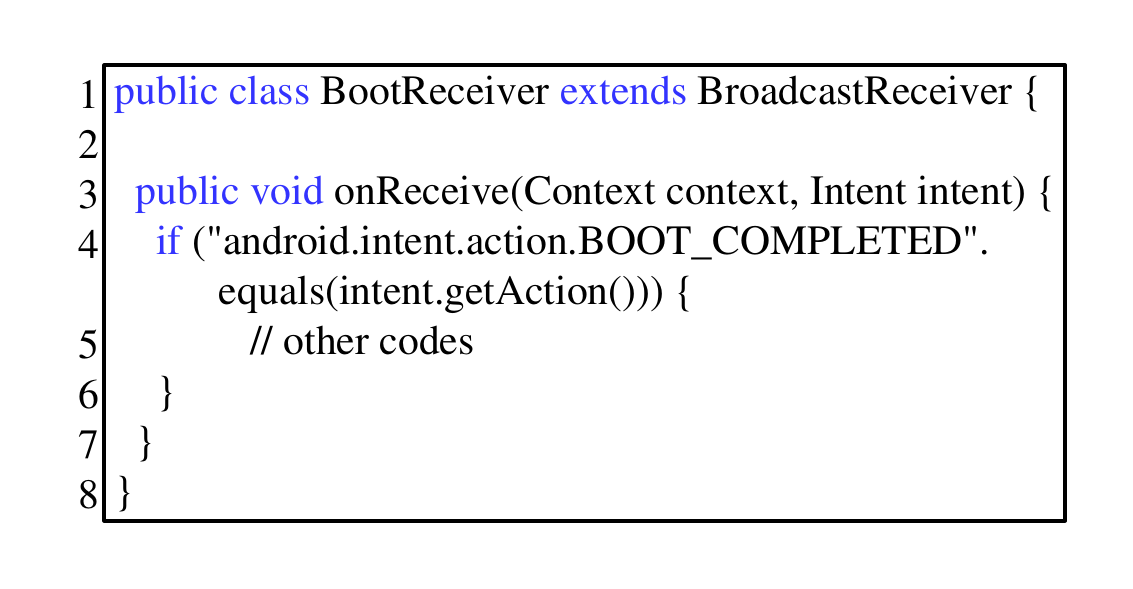}
  \vspace{-6ex}
  \caption{A typical code pattern for checking system-only broadcasts.}
  \label{fig:checkBroadcast}
\end{figure}

\end{document}